\documentclass[twocolumn]{aastex63}
\newcommand{\lya}{Ly$\alpha$}
\newcommand{\oiii}{[O\,{\sc iii}]}
\newcommand{\siii}{[S\,{\sc iii}]}
\newcommand{\hei}{He\,{\sc i}}
\newcommand{\ha}{H$\alpha$}
\newcommand{\hb}{H$\beta$}

\newcommand{\FeII}{Fe~{\sc ii}~}

\newcommand{\w}{$W_r$~}
\newcommand{\kms}{km s$^{-1}$}
\newcommand{\z}{$z$}

\newcommand{\MGII}{Mg~{\sc ii}}
\newcommand{\CIV}{C~{\sc iv}}
\newcommand{\myr}{$M_{\odot}$ yr$^{-1}$}

\received{}
\revised{}
\accepted{}
 
\shorttitle{MgII cold gas}
\shortauthors{Zou et al.}

\begin{document}

\title{Disturbed cold gas in galaxy and structure formation}
\author[0000-0002-3983-6484]{Siwei Zou}
\affiliation{Chinese Academy of Sciences South America Center for Astronomy, National Astronomical Observatories, CAS, Beijing 100101, China}
\affiliation{Departamento de Astronom\'ia, Universidad de Chile, Casilla 36-D, Santiago, Chile}

\author[0000-0003-3769-9559]{Robert A. Simcoe}
\affiliation{MIT Kavli Institute for Astrophysics and Space Research, 77 Massachusetts Avenue, Cambridge, MA 02139, USA}

\author{Patrick Petitjean}
\affiliation{Institut d'Astrophysique de Paris 98bis Boulevard Arago, 75014, Paris, France}

\author[0000-0002-4288-599X]{C\'eline P\'eroux}
\affiliation{European Southern Observatory, Karl-Schwarzschild-Str. 2, 85748 Garching near Munich, Germany}
\affiliation{Aix Marseille Univ., CNRS, LAM, (Laboratoire d'Astrophysique de Marseille), UMR 7326, F-13388 Marseille, France}


\author[0000-0002-6184-9097]{Jaclyn B. Champagne}
\affiliation{Steward Observatory, University of Arizona, 933 N Cherry Avenue, Tucson, AZ 85721, USA}

\author[0000-0002-7633-431X]{Feige Wang}
\affiliation{Department of Astronomy, University of Michigan, 1085 S. University Ave., Ann Arbor, MI 48109, USA}

\author[0000-0001-8405-2921]{Jinning Liang}
\affiliation{Department of Astronomy, School of Physics, Peking University, Beijing 100871, China}
\affiliation{Kavli Institute for Astronomy and Astrophysics, Peking University, Beijing 100871, China}

\author[0000-0001-6115-0633]{Fangzhou Jiang}
\affiliation{Kavli Institute for Astronomy and Astrophysics, Peking University, Beijing 100871, China}

\author[0000-0001-5951-459X]{Zihao Li}
\affiliation{Cosmic Dawn Center (DAWN), Denmark}
\affiliation{Niels Bohr Institute, University of Copenhagen, Jagtvej 128, DK-2200, Copenhagen N, Denmark}

\author[0000-0003-3995-4859]{Wen Sun}
\affil{Kavli Institute for Astronomy and Astrophysics, Peking University, Beijing 100871, China}
\affil{Department of Astronomy, School of Physics, Peking University, Beijing 100871, China}

\author[0000-0003-3310-0131]{Xiaohui Fan}
\affiliation{Steward Observatory, University of Arizona, 933 N Cherry Avenue, Tucson, AZ 85721, USA}

\author[0000-0001-5287-4242]{Jinyi Yang}
\affiliation{Department of Astronomy, University of Michigan, 1085 S. University Ave., Ann Arbor, MI 48109, USA}

\author[0000-0001-6947-5846]{Luis C. Ho}
\affil{Kavli Institute for Astronomy and Astrophysics, Peking University, Beijing 100871, China}
\affil{Department of Astronomy, School of Physics, Peking University, Beijing 100871, China}

\author[0000-0001-6052-4234]{Xiaojing Lin}
\affiliation{Department of Astronomy, Tsinghua University, Beijing 100084, China}
\affiliation{Steward Observatory, University of Arizona, 933 N Cherry Avenue, Tucson, AZ 85721, USA}

\author[0000-0002-1815-4839 ]{Jianan Li}
\affiliation{Department of Astronomy, Tsinghua University, Beijing 100084, China}

\author[0000-0002-6221-1829]{Jianwei Lyu}
\affiliation{Steward Observatory, University of Arizona, 933 N Cherry Avenue, Tucson, AZ 85721, USA}

\author[0000-0002-6540-7042]{Lile Wang}
\affiliation{Kavli Institute for Astronomy and Astrophysics, Peking University, Beijing 100871, China}

\author[0000-0002-7214-5976]{Weizhe Liu}
\affiliation{Steward Observatory, University of Arizona, 933 N Cherry Avenue, Tucson, AZ 85721, USA}

\author[0000-0002-6822-2254]{Emanuele Paolo Farina}
\affiliation{Gemini Observatory, NSF’s NOIRLab, 670 N A’ohoku Place, Hilo, HI 96720, USA}

\author[0000-0002-5768-738X]{Xiangyu Jin}
\affiliation{Steward Observatory, University of Arizona, 933 N Cherry Avenue, Tucson, AZ 85721, USA}

\author[0000-0003-0202-0534]{Cheng Cheng}
\affiliation{Chinese Academy of Sciences South America Center for Astronomy, National Astronomical Observatories, CAS, Beijing 100101, China}

\correspondingauthor{Siwei Zou}
\email{siwei1905@gmail.com}

\begin{abstract} 

Cold gas in the circumgalactic medium (CGM) and its interaction with galaxies remain poorly understood. Strong \MGII~($\lambda\lambda$2796, 2803) absorptions seen in background quasar spectra reveal large reservoirs of neutral hydrogen, potentially serving as progenitors of star-forming galaxies at high redshifts. In this study, we search for galaxies in the vicinity of very strong \MGII~absorbers (rest-frame equivalent width $W_r > 2$ \AA) with high kinematic velocities ($>$ 500 \kms) at $2.0 < z < 6.0$. Observations were conducted with VLT/MUSE, JWST/NIRCam, and ALMA to detect \lya~and nebular emission lines and dust continuum emission. We identify two \lya~emitters associated with a strong \MGII~absorber pair, separated by $\sim$1000 \kms~at $z\sim$ 4.87, in the vicinity of quasar J1306+0356. We observe relative differences in metallicity, dust content, and ionization states in this ultra-large absorption pair system, indicating potential metal and dust transfer within the system. For another strong \MGII~absorber at $z = 2.5662$ ($W_r = 2.638 \pm 0.124$ \AA) towards a second quasar J0305--3150, we detect a dusty star-forming galaxy at a projected distance of $D = $ 38 kpc. This galaxy exhibits prominent He~{\sc i}, [S~{\sc iii}], and Paschen$\gamma$ lines, along with significant dust continuum. It has a star formation rate of $\sim 121 \pm 33$ $M_{\odot}$/yr and likely harbors a rotating disk. These findings tentatively suggest that cold gas at high redshifts fuels disk formation and participates in metal and dust transfer within overdense CGM regions.

\end{abstract}


\section{Introduction}\label{sec_intro}

Cold and cool gas (T $\leq 10^4$ K) play a key role in galaxy formation and structural evolution. Simulations predict that cold gas flows into galaxies through cosmic filaments, further driving disk formation and galaxy evolution (see the review by \citealt{fag23} and references therein). The cold gas accretion mode in the circumgalactic medium (CGM), and its dependence on dark matter halo mass and redshift, remains incompletely characterized. In particular, resolving the kiloparsec-scale kinematics and dynamics of cold gas interacting with the disk, dust, and metals in high-redshift environments (especially at $z >$ 2) is still an open challenge. 

The Mg~{\sc ii} doublet ($\lambda\lambda$2796,2803) absorption is ubiquitously observed in the halos of galaxies, tracing cool (T $\sim$ 10$^4$ K) gas. Strong \MGII~absorbers (rest-frame equivalent width $W_r > 0.3$ \AA) are associated with optically thick neutral hydrogen (10$^{16.0}$ $<$ N (H~{\sc i}) $<$ 10$^{22.0}$ cm$^{-2}$) and, consequently, with star forming activities \citep{bar14}. The evolution of strong Mg~{\sc ii} absorber mirrors the global cosmic star formation rate at 2 $< z <$ 6 \citep{chen17,zou21,seb24}. These absorbers are found in association with star-forming galaxies \citep{men11,lan18,not12b,zou24a} and metal-enriched overdense regions \citep{dutta20,wu23,zou24a,zou24b}.

The ultra-strong Mg~{\sc ii} systems (hereafter USMg~{\sc ii}, $W_r (\lambda2796)> 3$ \AA) trace cold and dense gas, but their origins remain debated. Kinematically broad \MGII~systems have been observed to be associated with highly disturbed cold (and even molecular) gas \citep{zou18}. Two primary scenarios have  proposed to explain the origin of ultra-strong Mg~{\sc ii} absorbers: (a) they may arise from starburst-driven outflows or tidally-stripped gas resulting from a major interaction \citep{nes11}. In this scenario, the $W_r$ and velocity spread ($\Delta v$) of the gas correlate with the star formation rate (SFR) and luminosity of the host galaxy \citep{kac10}; (b) multiple `normal' galaxies within a galaxy group could contribute to the strong absorption strength and kinematics observed \citep{kac10,rubin11,rubin14}. Both scenarios have observational support at $z < 1$ \citep{nes07,gau13,guha22,guha24}. \citet{nes07} and \citet{pro06} studied USMg~{\sc ii} systems at $z < 1$ and identified bright galaxies at relatively low impact parameters to the absorption, supporting the outflow model. Additionally, \citet{gau13} reported a USMg~{\sc ii} system at $z \sim 0.5$ associated with three galaxies within 200 kpc, suggesting the possibility of the galaxy group scenario. It is also plausible that these two mechanisms coexist.


The link between the physical properties of the multiphase CGM and the surrounding environment has been explored in several surveys using MUSE, ALMA, and JWST. The MUSE-ALMA Haloes Survey \citep{per22} combined data from VLT/MUSE, ALMA, and HST to investigate strong H~{\sc i} and metal-absorbing gas at $0.3 < z < 1.4$. The study revealed that damped Ly$\alpha$ systems (DLAs, log N(H~{\sc i}) $>$ 20.3) with strong \MGII~absorption at $z < 1$ are associated with multiple star-forming galaxies. The kinematics of the absorbing gas are linked to the ionized and molecular gas in the surrounding galaxies \citep{per19,sza21,weng23}. The MusE GAs FLOw and Wind (MEGAFLOW) survey has identified a bimodal relationship between the azimuthal angle and metallicity of cool gas at $z \sim 1$ \citep{wen21}, suggesting that \MGII~gas traces either galactic inflow \citep{zab19} or galactic outflow \citep{sch19}. The MUSE Analysis of Gas around Galaxies (MAGG) survey probes the connection between [O~{\sc ii}] emission and \MGII~absorption at $z \sim$ 1 \citep{dutta21}, and between \lya~emission and C~{\sc iv} and Si~{\sc iv} absorbers at $z$ = 3--4.5 \citep{magg_lofthouse,marta24}, finding that metal-enriched absorbing gas strength and kinematics correlate positively with the number of \lya~emitters (LAEs). Additionally, ALMA observations indicate that DLA systems at $z \sim 4$ are associated with [C~{\sc ii}] emission in disk galaxies. \citet{prochaska19} and \citet{neeleman19} reported the detection of [C~{\sc ii}] emission from five out of six high-metallicity ([X/H] $>$ --1.36) DLAs at $z \sim 4$, suggesting a connection between the velocity of [C~{\sc ii}] emission and DLAs at this redshift.

Recent JWST observations have identified other nebular emission lines, including \ha, \oiii~and \hb~near \MGII~absorbers at $z$ = 2--7 \citep{bordoloi23,wu23,christensen23,zou24b}. However, the majority of \MGII~absorbers in \citet{bordoloi23} are classified as weak \MGII~systems.

\begin{figure*}[ht]
    \centering
    \resizebox{0.5\hsize}{!}{\includegraphics[width=\textwidth]{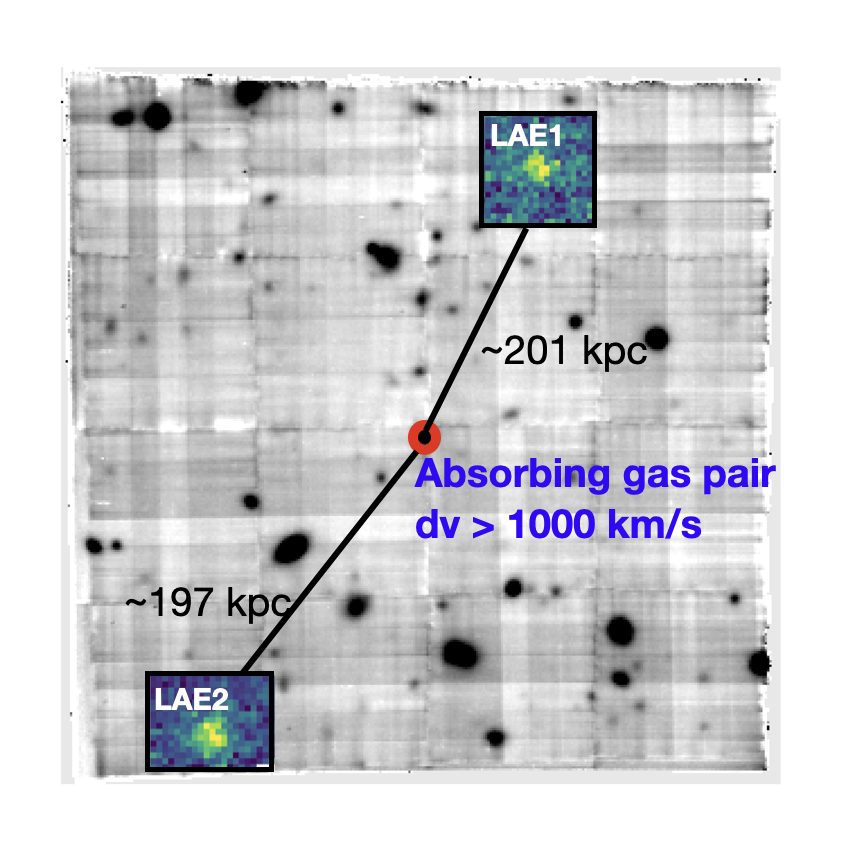}}
    \resizebox{0.49\hsize}{!}{\includegraphics[width=\textwidth]{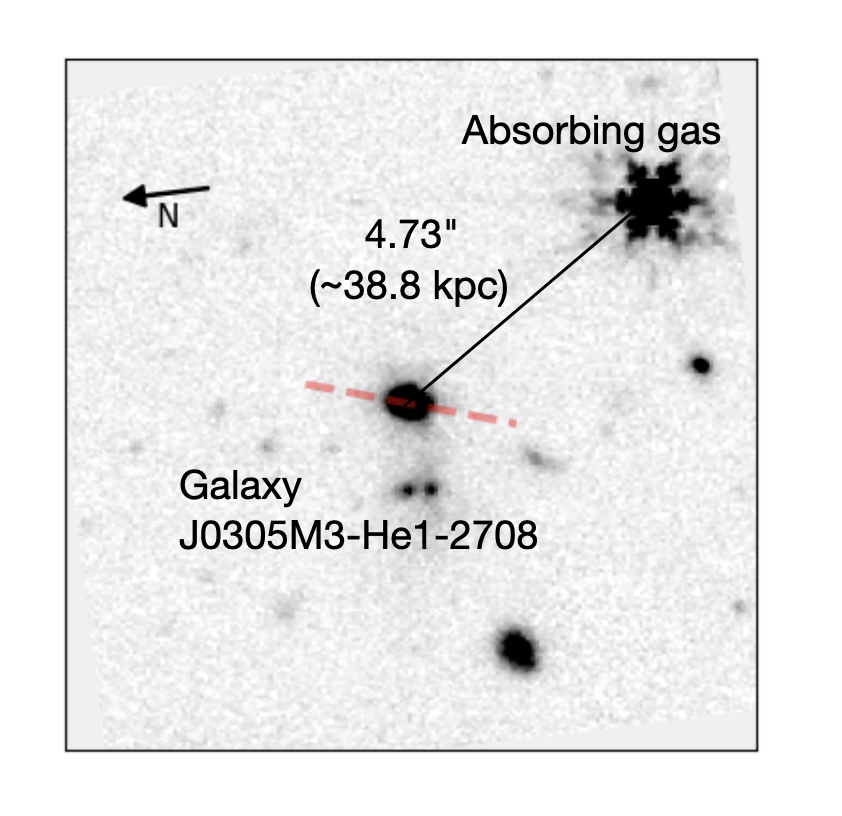}}
    \caption{$Left.$ Spatial distribution of the absorbing gas along J1306+0356 (located at the center of the VLT/MUSE cube) and the two LAEs, which are offset by $\sim 200$ kpc from the gas. The background image is the variance-weighted white-light image generated by \textsc{\textsc{MUSELET}} (see details in Section~\ref{sec:emission_line}). Narrow-band images of the two detected LAEs are presented. The absorbing gas structure is composed of two VSMGII systems, with a total velocity width $\Delta v$ greater than 1000 \kms. $Right.$ Spatial distribution of the absorbing gas along the J0305--3150 sightline (upper right corner) and the host galaxy, as detected with JWST/NIRCam data. The red dashed line is the fitted galaxy major axis.}
    \label{fig:twosystems_spatial}
\end{figure*}

To study the disturbed cold gas in the CGM and its interaction with galaxies and large-scale structures at $z > 2$, we conducted an initial sample. We first checked all sightlines with strong \MGII~absorbers that have archival VLT/MUSE coverage, and then identify those that are also covered by JWST and ALMA data. 
To date, no dedicated survey has focused on studying the galaxy environments around very strong and kinematically broad Mg~{\sc ii} absorbers at $z > 2$. In this paper, we report the discovery of \lya~emission at $z = 4.86$ and He~{\sc i}, [S~{\sc iii}], Paschen lines, and dust continuum at $z = 2.56$ around three very strong ($W_r$ $> 2.5$ \AA (see Figure \ref{fig:twosystems_spatial}), we refer to very strong Mg~{\sc ii} systems, VSMGII, throughout the paper) and kinematically complex ($\Delta v > 500$ \kms) \MGII~systems. 

The paper is structured as follows: We present the sample selection and data reduction in Section \ref{sec:sample_selection}. We detail the detection of the \lya~emitters around the $z \sim 4.9$ absorbing system and the dusty galaxy around the absorber at $z \sim 2.6$ in Section \ref{sec:results}. Section \ref{sec:discussion} discusses the interpretation of these systems. Finally, we summarize our findings in Section \ref{sec:summary}. Throughout this work, we adopt the standard $\Lambda$CDM model with the following cosmological parameters: $H_0 = 70$ km s$^{-1}$ Mpc$^{-1}$, $\Omega_\Lambda = 0.69$, and $\Omega_m = 0.31$.

\section{Sample selection and data reduction}{\label{sec:sample_selection}}

\subsection{Sample selection and detection limits}

The parent sample of the Mg~{\sc ii} absorber catalogs is from \citet{chen17} and \citet{zou21}. \citet{chen17} comprises the Magellan/FIRE sample of 100 quasars with emission redshifts between $z = 3.55$ and $z = 7.09$, with an average spectral resolution of $R \sim 6000$ \citep{mat13}. \citet{zou21} includes Gemini near-infrared spectroscopy of 51 quasars at $z > 5.7$, with an average spectral resolution of $R \sim 800$ \citep{shen19}. Considering the uncertainties in the equivalent width measurements, we pre-select the initial sample based on the criteria of $W_r > 2.0$ \AA~and $\Delta v > 500$ \kms, rather than $W_r > 3.0$ \AA, to avoid missing potential targets. After this step, we checked ancillary data from three sources: archival VLT/MUSE data, the JWST GO1 survey A Spectroscopic Survey of Biased Halos in the Reionization Era (ASPIRE), and ASPIRE-ALMA data. 

We list six sightlines with VSMGII absorbers and archival VLT/MUSE data covered in Table \ref{table:qso_info}. Except for the deepest 10-hour exposure datacube in the field of J1306+0356, we do not detect \lya~emission in the other five MUSE datacubes. The flux detection limits for the 10-hour MUSE-Deep single-point data and 1-hour MUSE-Wide data are $3.1 \times 10^{-19}$ erg s$^{-1}$ cm$^{-2}$ \citep{bacon17} and $2.0 \times 10^{-17}$ erg s$^{-1}$ cm$^{-2}$ \citep{muse-wide}, respectively.  For JWST/NIRCam WFSS data in the F356W band, the 5$\sigma$ line detection limit can reach $2.0 \times 10^{-18}$ erg s$^{-1}$ cm$^{-2}$ by integrating the spectra over 50~\AA~\citep{wang23}, which corresponds to a luminosity limit of $1.1 \times 10^{40}$ erg s$^{-1}$ at $z = 2.5$. Given that the second strong \MGII~system toward J1306+0356 (denoted as $b$ in Table \ref{table:qso_info}) is not listed in \cite{chen17}, we re-measure the equivalent width and velocity spread of the absorbing systems along the two sightlines analyzed in this work. We do not re-measure the remaining four sightlines listed in Table \ref{table:qso_info}, as no associated galaxies are detected and no JWST or ALMA data are available.

\begin{table*}
\centering
  \caption{Quasar and absorber information in the sample. We list the targets that have MUSE data. We remeasure the absorber rest-frame equivalent width and velocity width for the systems towards J0305--3150 and J1306+0356. Other values are presented in \citet{chen17}. The labels ``sys1'' and ``sys2'' for J1306+0356 indicate the two metal-absorbing systems shown in Figure~\ref{fig:J1306_abs_complete}. \label{table:qso_info}}
\begin{tabular}{lllcclll}
\hline
QSO           & RA          &  DEC         & $z_{abs}$ & $W_r$($\lambda$2796)  & $\Delta v$ & Exposure Time  & PI \& Project ID   \\
              &             &              &      & (\AA) &\kms    &   (hours)  & \\
\hline
{\bf J0305--3150} & 03:05:16.9  & --31:50:55.98 & 2.5662 & 2.869$\pm$0.481 & 524.6 & 2.5 &   Venemans 094.B-0893 \\
J0836+0054 & 08:36:43.9  & +00:54:53.3   &  3.7443 & 2.509$\pm$0.016 & 510.4 &  5.3 &  Bian  106.210Z\\  
{\bf J1306+0356$_\textrm{sys1}$} & 13:06:08.3  & +03:56:26.3   &  4.8634 & 2.891$\pm$0.109  & 541.5 & 10.4         &  Ellis 0103.A-0140  \\ 
{\bf J1306+0356$_\textrm{sys2}$} & 13:06:08.3  & +03:56:26.3   &  4.8821 & $<$ 2.959 &  465.9 & 10.4         &  Ellis 0103.A-0140  \\ 
J1509-1749 & 15:09:41.78 & --17:49:26.80 &  3.3925 & 5.679$\pm$0.056 & 811.5 & 1.0 & Farina 0101.A-0656 \\ 
J2211-3206 & 22:11:12.39 & --32:06:12.95 &  3.7144 & 3.505$\pm$0.068 & 623.4 & 1.0      &   Farina 0101.A-0656 \\ 
J1030+0525 & 10:30:27.09 & +05:24:55.02  &  2.78  & 2.617$\pm$0.069   & 583.9  & 1.0       &  Karman 095.A-0714 \\
\hline
\end{tabular}
\end{table*}


\subsection{MUSE data reduction}

The MUSE data used to search for \lya~emitters in the quasar field of J1306+0356 is collected from the MUSE-DEEP survey (P.I. R. Ellis, Program ID: 0103.A-0140). We used the ESO MUSE data reduction pipeline \citep{bacon17,wei20}\footnote{https://www.eso.org/sci/software/pipelines/muse/} to reduce the datacube. The ESO pipeline reduction process is briefly described as follows: First, we process the raw datacube using the ESO MUSE pipeline v2.8.9. Daily calibrations and the geometry table are used to generate pixel information, including location, wavelength, photon count, and variance estimates. Each exposure was precisely recentered to correct for derotator wobble. We then use the $scipost$ recipe to perform astrometric and flux calibrations on the pixtable. The final datacube was generated using the $makecube$ recipe. Sky background subtraction was performed using the ZAP process, as described in \citet{soto16}.



For the inter-stack masking, we run the $scibasic$ and $scipost$ recipes twice: the first time without using the specific bad pixel table, and the second time with it. Using the output of the `bad-pixel' version of the cube, we derive a new, 3D mask which we apply to the original cube, effectively removing the inter-stack bad data.

\subsection{JWST data reduction}

The JWST data is taken from the A SPectroscopic survey of biased halos In the Reionization Era (ASPIRE) JWST GO1 survey (P.I. F. Wang, Program ID: \#2078). The science goal of this program is driven by the study of supermassive black hole growth, its connection with the host galaxy, and metal enrichment during the reionization epoch. The full data JWST/NIRCam WFSS data reduction is described in \citet{wang23} and \citet{yang23}. We perform data using version 1.10.2 \citep{jwst_pipeline} of the JWST Calibration Pipeline (CALWEBB). For data calibration, we employ reference files (jwst 1080.pmap) from version 11.16.21 of the standard Calibration Reference Data System (CRDS). All imaging data in this study align with the Gaia data release 3. We adopt the method outlined in \citet{fengwu22} which includes a column-average correction for 1/f-noise. This model traces point sources observed in the Large Magellanic Cloud field.

\subsection{ALMA data reduction}
The ALMA data used in this work is from the Cycle 9 large program (P.I. F. Wang, Program ID: 2022.1.01077.L) for the 25 ASPIRE quasar fields with ALMA band-6 mosaics. The data reduction was processed with the CASA pipeline v6.4.1.12. The full details of the ALMA data reduction are presented in \citet{fengwu24}. For the continuum imaging, we flagged all spectral channels with a velocity offset smaller than 500 \kms~from the quasar [C~{\sc ii}] line center. Noisy spectral channels caused by telluric absorption were also visually identified through \texttt{plotms} and flagged from the continuum imaging. We then split the measurement sets to a channel width of 125 MHz and obtained continuum imaging at native resolution (robust=0.5, no uv-tapering; typical beam size is about 0.7$\arcsec$) and tapered resolution (robust=2.0, uv-tapered with a Gaussian kernel of full-width-half-maximum (FWHM) of 1.0$\arcsec$).

\subsection{HST archived data}
For sightline J1306+0356, we used archival Hubble Space Telescope (HST) imaging in the F775W and F850LP bands (PI: M. Stiavelli; Program ID \#9777) to estimate the LAE UV continuum. For sightline J0305-3150, we used archival HST data, including WFC3 imaging (F105W, F125W, F160W) and ACS imaging (F606W, F814W) (PI: C. Casey; Program ID \#15064), for photometric measurements and the spectral energy distribution (SED) fitting discussed in Section \ref{sec:sed}.


\subsection{Emission lines search}\label{sec:emission_line}
We search for \lya~emission in the MUSE field using a simple SExtractor-based Python tool, \textsc{MUSELET}\footnote{https://mpdaf.readthedocs.io/en/latest/muselet.html}. \textsc{MUSELET} uses SExtractor\footnote{http://www.astromatic.net/software/sextractor}\citep{ber96} to search for emission lines in narrowband (nb) imaging. The process is as follows: First, \textsc{MUSELET} creates a variance-weighted white light image and RGB images based on one-third of the wavelength range each. Next, \textsc{MUSELET} runs SExtractor on the nb files and generates a catalog that distinguishes sources with continuum emission. The continuum is extracted from a range of 20 wavelength pixels around the \lya~emission line. We then visually inspect the catalog and rule out lines at lower redshifts. The final detected two LAEs are presented in Figure \ref{fig:lya}, both exhibiting a double-peaked \lya~profile.

For the VSMGII at $z =$ 2.5662 along the quasar sightline J0305--3150, we searched for the [S~{\sc iii}]($\lambda$9531), He~{\sc i}($\lambda$10830), and Paschen$\gamma$ lines using a method similar to that used for the [O~{\sc iii}] and H{$\beta$} lines in \citet{wang23}, Section 3. We identified flux peaks with S/N $>$ 3 from the 1D spectra and applied a Gaussian filter template for the He~{\sc i} and [S~{\sc iii}] lines to select potential candidates. The selected targets were then visually inspected by more than three people to determine the final set of emission-line galaxies.




\section{Results}\label{sec:results}

\subsection{Absorption profile}
The VSMGII detected at $z_{\mathrm{abs}} = 4.8651$ towards quasar J1306+0356 ($z_{\mathrm{em}} = 6.02$, R.A. = 13$^{\mathrm{h}}$06$^{\mathrm{m}}$08.27$^{\mathrm{s}}$, Dec. = +03$^\circ$56$'$26.36$''$) is reported with $W_r$ = 2.804 \AA~and a velocity spread of $\Delta v$ = 180.8 \kms~in \citet{chen17}. Al~{\sc ii} and Fe~{\sc ii} absorption is reported in the same system in \citet{jiang07}. After our re-measurements of the $W_r$ and $\Delta v$ (see Table \ref{table:qso_info}), we report a new system at $z$ = 4.8821, which is offset by around {\bf 1000 \kms~}from the system at $z = 4.8651$. The second system has an $W_r$(\MGII($\lambda$2803)) = 2.371 $\pm$ 0.097 \AA~and $\Delta v$ = 421.9 \kms. We redefine the centers of these two systems at the positions of the strongest \MGII($\lambda$2796) subcomponents, with $z = 4.8634$ and $z = 4.8821$, respectively. 

\begin{figure*}
  \resizebox{0.9\hsize}{!}{\includegraphics{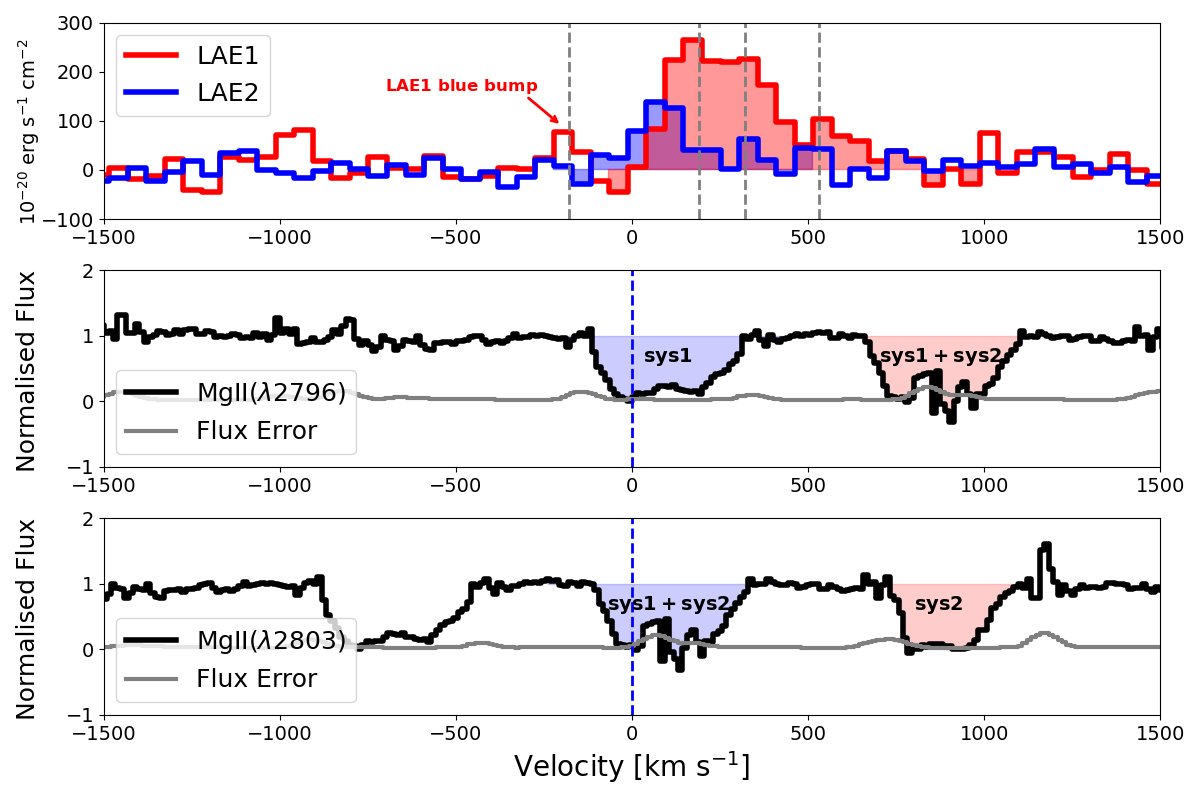}}
  \caption{\small{Emission profile of the \lya~lines detected with VLT/MUSE and the absorption profile of the \MGII~absorber observed by Magellan/FIRE along the J1306+0356 sightline at $z = 4.8651$, shown in velocity space. The upper panel shows the spectra of the two LAEs, which are at distances of 201 kpc (red) and 197 kpc (blue) from the absorbing gas, respectively. LAE1 is centered at $z = 4.867$ at the peak of the strongest component. The grey dashed lines indicate the potential subcomponents of LAE~1. The shaded regions indicate the wavelength interval used to compute the equivalent width (EW) of the \lya~lines. For LAE1, the blue bump is tentative, therefore, we do not include it in the EW measurement. The lower panels show the \MGII~$\lambda2796$ and \MGII~$\lambda2803$ lines, respectively, with the blue dashed line representing the zero point at $z = 4.8651$, where the first major subcomponents are located. Complete Voigt profile fitting of the two systems is provided in Table~\ref{table:absorption} and Appendix Figure~\ref{fig:J1306_abs_complete}.
}}\label{fig:lya}
\end{figure*}

We present the two absorbing systems and detected two LAEs in Figure \ref{fig:lya}, with details in Figure \ref{fig:J1306_abs_complete}. 
In addition to \MGII, we detect \FeII, Mg~{\sc i}, Si~{\sc ii}, Al~{\sc ii}, Al~{\sc iii}, C~{\sc iv}, and tentative detections of Zn~{\sc ii} absorption lines. We denote the system at $z = 4.8634$ as sys 1 and the system at $z = 4.8821$ as sys 2. {\bf For \MGII, the \MGII($\lambda$2803) line in sys 1 is blended with the \MGII($\lambda$2796) line in sys 2.} In Table \ref{table:absorption}, we present the $W_r$ of system 1 \MGII($\lambda$2796) line, and the blended region. For \CIV~lines, there is also possible blending between C~{\sc iv}($\lambda$1550) in system 1 and C~{\sc iv}($\lambda$1548) in system 2. We use the red and blue shaded regions in Figure \ref{fig:J1306_abs_complete} to present this absorbing system pair including the blended region. 
Given that the \MGII~and \CIV~lines are likely strongly saturated, we only fit the Si~{\sc ii}, Al~{\sc ii}, and Al~{\sc iii} lines with Voigt profiles including multiple subcomponents using VoigtFit \citep{kro18}. We adopted Doppler parameters $b$ in the range 5–30 \kms, and present in Table \ref{table:absorption} the best-fit results corresponding to the minimum $\chi^2$, obtained with $b = 25$ \kms.

From Table \ref{table:absorption}, we observe that in general, system 2 has higher $W_r$ for most ions, except for Al~{\sc iii}, \CIV, and slightly weaker Mg~{\sc i} lines. Given that system 1 has weaker intermediate ions but stronger high-ionization lines, it appears to be more ionized and less dusty than system 2. We present the CLOUDY modelling of the ionization state and related discussion of these two systems in Section \ref{sec:cloudy}.

In Figure~\ref{fig:J0305_emi_abs}, we present the very strong Mg~{\sc ii} system detected at $z = 2.5662$, alongside the Fe~{\sc ii}~($\lambda2600$) line. Other UV lines, such as O~{\sc i} and Si~{\sc ii}, are shifted into the optical spectral range and are absorbed by the IGM because the quasar is at $z > 5$. The Mg~{\sc ii} doublet profiles show three major subcomponents.

\subsection{Galaxy properties around the very strong \MGII~systems}

\subsubsection{\lya~emitters around the \MGII~absorption pair at $z\sim$ 4.86}

Using \textsc{MUSELET}, we obtain the initial list of \lya~candidates in the J1306+0356 field. We then visually inspect the candidates by examining their 2D and 1D spectra, confirming two LAEs: LAE1 at RA, Dec = 196.5397, 3.9328; and LAE2 at RA, Dec = 196.5312, 3.9483. The spectra of the two LAEs and the VSMGII absorber are presented in Figure \ref{fig:lya}. The $v = 0$~\kms\ corresponds to $z = 4.8634$, at the strongest component of LAE1. Note that the redshift here is not the systematic redshift ($z_{\textrm{sys}}$), given that there is often a velocity shift between the \lya~line peak and $z_{\textrm{sys}}$ in previously reported LAEs (e.g., \citealt{hashimoto17}). We calculate the $z_{\textrm{sys}}$ uncertainties in Table~\ref{table:emi_info} based on its relation with the red (peak) component and the blue bump velocity differences in \citet{ver18}.


We then measure the \lya~equivalent width (EW) of the \lya~emission. The EW of the \lya~line is defined as the ratio of the \lya~line luminosity to the UV continuum luminosity density at \lya = 1215.67 \AA~\citep{sch03}:


\begin{equation}\label{eq:ew}
EW = \int_{\lambda_{2}}^{\lambda_{1}}
\frac{f_{\mathrm{Ly}\,\alpha}^{\mathrm{line}} -
      f_{\mathrm{Ly}\,\alpha}^{\mathrm{cont}}}
     {f_{\mathrm{Ly}\,\alpha}^{\mathrm{cont}}}
\, d\lambda
\end{equation}

\noindent where $f_{\text{Ly}\alpha}^{\text{line}}$ and $f_{\text{Ly}\alpha}^{\text{cont}}$ are the flux densities of the \lya~line and the continuum, respectively. We measure the \lya~line flux from the extracted 1D spectra, including the blue bump. 
We then use archival Hubble Space Telescope (HST) F775W and F850LP data to estimate the UV continuum magnitude and flux density upper limit. From this, we calculate the EW of LAE1 and LAE2 to be {\bf 73.15 $\pm$ 24.05 \AA} and {\bf 20.01 $\pm$ 9.27 \AA}, respectively. 
We corrected for instrumental broadening using the MUSE line-spread function when calculating the FWHM of the emission line. We estimate the star formation rate (SFR) of the two LAEs from the \lya~luminosity, using the SFR-H$\alpha$ relation in \citet{ken98} and the \lya-H$\alpha$ emissivity ratio in \citet{oster06}: SFR(\lya) = 9.1 $\times$ 10$^{-43}\times$ L(\lya) $M_\odot$/yr, yielding SFR(\lya) = 5.75 $\pm$ 1.89 $M_\odot$/yr and 1.55 $\pm$ 0.73 $M_\odot$/yr, respectively. This SFR value is consistent with the results in \citet{sob19}, which measure SFR(\lya) at $z > 2.1$. Details of these two LAE measurement are presented in Table \ref{table:emi_info}.


\begin{table}[h]
\centering \caption{Galaxy property measurements around the VSMGII absorbers}
\begin{tabular*}{\columnwidth}{@{\extracolsep{\fill}}ll}
\hline
\multicolumn{2}{c}{\bf J1306+0356 LAE1} \\ 
\hline
$z_{\textrm{Ly}\alpha}$    &  4.867 $\pm$ 0.004 \\
R.A.        &  13:06:09.53   \\
Dec.        &  +03:55:58.08   \\
$D$ (kpc)   &  201 \\ 
$f_{\textrm{Ly}\alpha}$ (10$^{-17}~$erg s$^{-1}$ cm$^{-2}$)  &   2.45 $\pm$ 0.81   \\
$L_{\textrm{Ly}\alpha}$ (10$^{42}$ erg s$^{-1}$) & 6.32 $\pm$ 2.08 \\
EW (\AA)    &   73.15 $\pm$ 24.05  \\
FWHM (\kms) &  121 \\
SFR (\myr)  &  5.75 $\pm$ 1.89  \\ 
\hline
\multicolumn{2}{c}{\bf J1306+0356 LAE2} \\ 
\hline
$z_{\textrm{Ly}\alpha}$    &  4.865 $\pm$ 0.004 \\
R.A.        & 13:06:07.49  \\
Dec.        & +03:56:53.88  \\
$D$ (kpc)   &   197 \\ 
$f_{\textrm{Ly}\alpha}$ (10$^{-17}~$erg s$^{-1}$ cm$^{-2}$)  &  0.67 $\pm$ 0.31   \\
$L_{\textrm{Ly}\alpha}$ (10$^{42}$ erg s$^{-1}$) & 1.70 $\pm$ 0.80 \\
EW (\AA)    &   20.01 $\pm$ 9.37  \\
SFR (\myr)  &  1.55 $\pm$ 0.73 \\
\hline
\multicolumn{2}{c}{\bf J0305M3150-HeI-2708} \\ 
\hline
$z_{em}$    &  2.564 \\
R.A.        &  03:05:17.11     \\
Dec.        &  --31:50:52.08   \\
$D$ (kpc)   &  38 \\ 
$f_{\textrm{He~{\sc i}}\lambda 10830}$ (10$^{-17}~$erg s$^{-1}$ cm$^{-2}$)& 1.37 $\pm$ 0.42  \\
FWHM (\kms) &  710 \\
$f_{\textrm{[S~{\sc iii}]}\lambda 9531}$ (10$^{-17}~$erg s$^{-1}$ cm$^{-2}$)& 1.22 $\pm$ 0.52 \\
$f_{\textrm{[S~{\sc iii}]}\lambda 9067}$ (10$^{-17}~$erg s$^{-1}$ cm$^{-2}$)& 1.01 $\pm$ 0.49   \\
$f_{\textrm{Pa$\gamma$}}$ (10$^{-17}~$erg s$^{-1}$ cm$^{-2}$)  & 1.07 $\pm$ 0.38 \\
$f_{\textrm{Pa10}}$ (10$^{-17}~$erg s$^{-1}$ cm$^{-2}$)  &  0.89 $\pm$ 0.43 \\
EW$_{\textrm{He~{\sc i}}\lambda 10830}$ (\AA)  & 329.51 $\pm$ 51.0 \\
EW$_{\textrm{[S~{\sc iii}]}\lambda 9531}$ (\AA)  &  275.31 $\pm$ 66.16  \\
EW$_{\textrm{Pa$\gamma$}}$ (\AA)  &  124.19 $\pm$ 44.11   \\
$S_{\textrm{cont}}$ (mJy) &  424 $\pm$ 110 \\
$L_{\textrm{FIR}}$ (erg s$^{-1}$)  & (5.74 $\pm$ 1.58) $\times 10^{11}$ \\ 
log $M_{\textrm{*}}$/$M_{\odot}$  & 10.73 $^{+0.02}_{-0.60}$  \\ 
log $M_{\textrm{dust}}$/$M_{\odot}$  & 8.30 $\pm$ 0.11  \\ 
SFR (\myr)  &  121 $\pm$ 33   \\ 
\hline
\end{tabular*}
\label{table:emi_info}
\end{table}



\subsubsection{Dusty star-forming galaxy around the VSMGII absorber at $z =$ 2.5662 in the J0305-3150 field}\label{sec:sfg_absorber}

\begin{figure*}
\resizebox{\hsize}{!}{\includegraphics{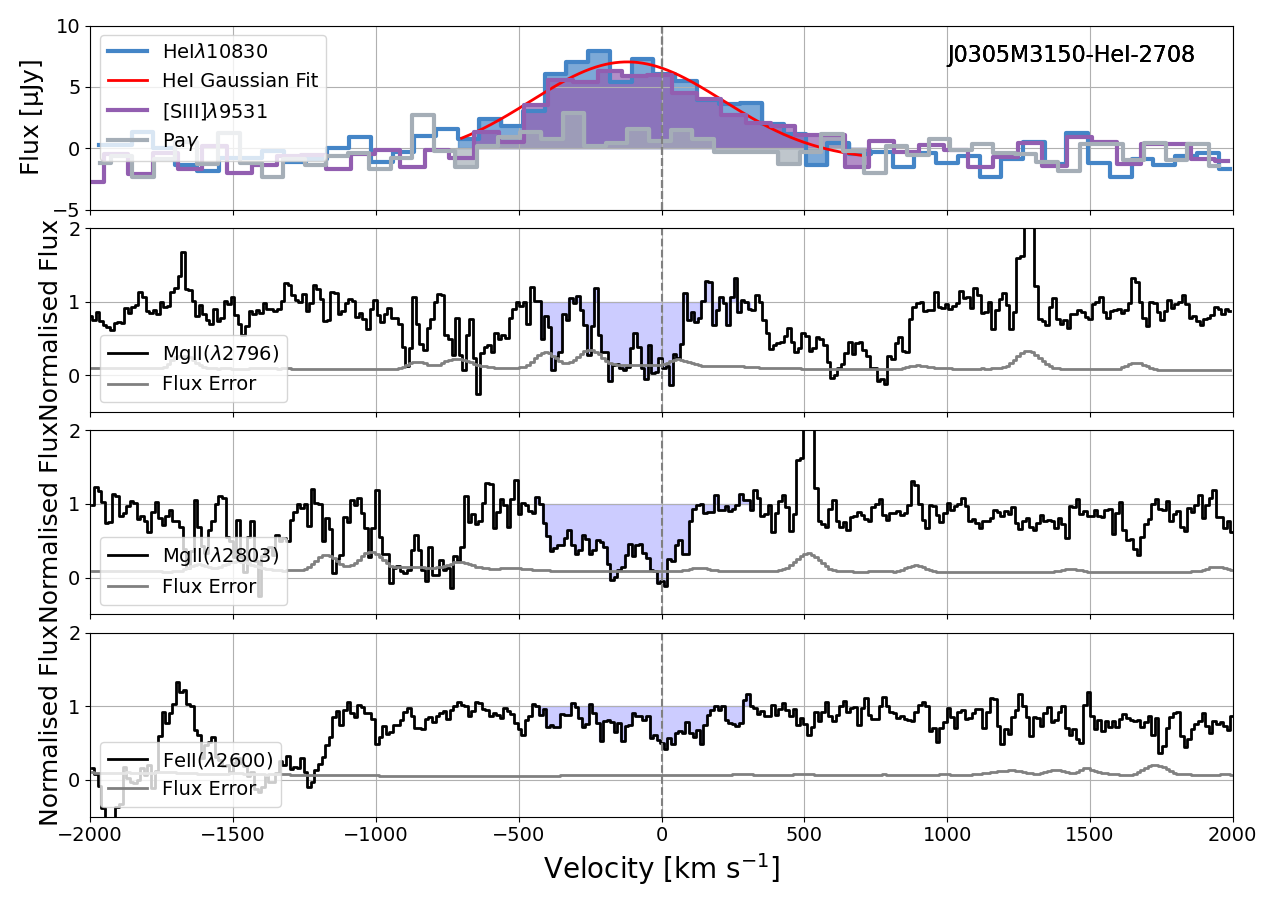}}
  \caption{\small{
  The absorption profiles of the VSMGII at $z$ = 2.5662, along with the 1D spectra of the detected galaxy which is 4.65$\arcsec$ (38 kpc) away from the absorber. The galaxy shows \hei (with Gaussian fit), \siii~and Paschen emission lines at $z$ = 2.5643 in the JWST NIRCam F356W band. The zero point is centered on the strongest \MGII~subcomponent (the same in each panel). 
}}\label{fig:J0305_emi_abs}
\end{figure*}

We detect a galaxy counterpart ASPIRE J0305M31-He1-2708 (RA, DEC = 03:05:17.09, --31:50:49.20) in the J0305--3150 field with prominent He~{\sc i}($\lambda$10830), [S~{\sc iii}]($\lambda$9531), [S~{\sc iii}]($\lambda$9067), Pa$\gamma$ and Pa10 lines. Profiles of the absorbing gas and galaxy are presented in Figure \ref{fig:J0305_emi_abs}. Full details of the galaxy’s 2D/1D spectra and emission lines are shown in Figure \ref{fig:J0305_emi} and Table \ref{table:emi_info}. The FWHM of He~{\sc i} line is around 710 \kms, with instrument broadening corrected using NIRCam Grism LSF function \citep{greene17}. The flux ratio between He~{\sc i} and Pa$\gamma$ line is 2.7.

We measure the $W_r$ of the emission lines with the F115W, F200W and F356W photometric data and the line flux. The target continuum flux densities is modelled with a power-law $f_{cont} = \lambda^\alpha$. Then we estimate the flux densities at the emission lines and measure the $W_r$ following same method in Equation \ref{eq:ew}. 

\subsubsection{Stellar mass, Dust Continuum and SFR}\label{sec:sed}

We also detected dust continuum in the galaxy J0305M50-HeI-2708 using ASPIRE-ALMA data (noted as J0305M50.C04, see Figure \ref{fig:J0305_jwst_alma}). The details of searching dust continuum and [C~{\sc ii}] emitters at $z > 4$ using the ASPIRE-ALMA data is presented in \citet{fengwu24}. 

The aperture and peak fluxes of the J0305M3150.C04 source are 0.42 $\pm$ 0.11 mJy and 0.40 $\pm$ 0.03 mJy, respectively. We calculate the far-infrared SFR following the relation from \citet{mur11}: SFR = 2.1 $\times$ 10$^{-10}$ $L_\textrm{FIR(SF)}$ [\myr], where $L_\textrm{FIR}$ is the far-infrared luminosity. The $L_\textrm{FIR}$ is determined by SED fitting of a black body in the 42.5--122.5 $\mu$m range. We adopt an average dust temperature of 35 K and a dust emissivity index $\beta$ of 1.6 \citep{neeleman18}. We then calculate the dust mass and $L_\textrm{FIR(SF)}$, which are found to be (1.98 $\pm$ 0.55) $\times$10$^{8} M_{\odot}$ and (5.74 $\pm$ 1.58) $\times$10$^{11}$ erg s$^{-1}$, respectively. The resulting SFR is 121$\pm$ 33 \myr, indicating intense star formation in this galaxy.

The stellar mass of the galaxy is obtained using the spectral energy distribution fitting tool BAGPIPES \citep{car18}. We use multiband photometry from JWST (F115W, F200W, F356W) and HST (F105W, F125W, F160W, F606W, and F814W bands) as inputs. We adopt a constant star formation history (SFH) and the Calzetti dust law \citep{cal00}, allowing $A_V$ to vary between 1.0 and 5.0, metallicity (log $Z/Z_\odot$) to vary between $-2$ and $2$, and age to vary between 100 Myr and 2.5 Gyr. The best-fitted stellar mass of the galaxy is $\log(M/M_{\odot}) = 10.73^{+0.02}_{-0.60}$ (see Figure \ref{fig:J0305_sed}). The difference in the stellar mass varies by around 0.2 dex when adopting a post-starburst SFH.

\subsubsection{Galaxy morphology}
\begin{figure*}
\centering
\includegraphics[width=0.99\textwidth]{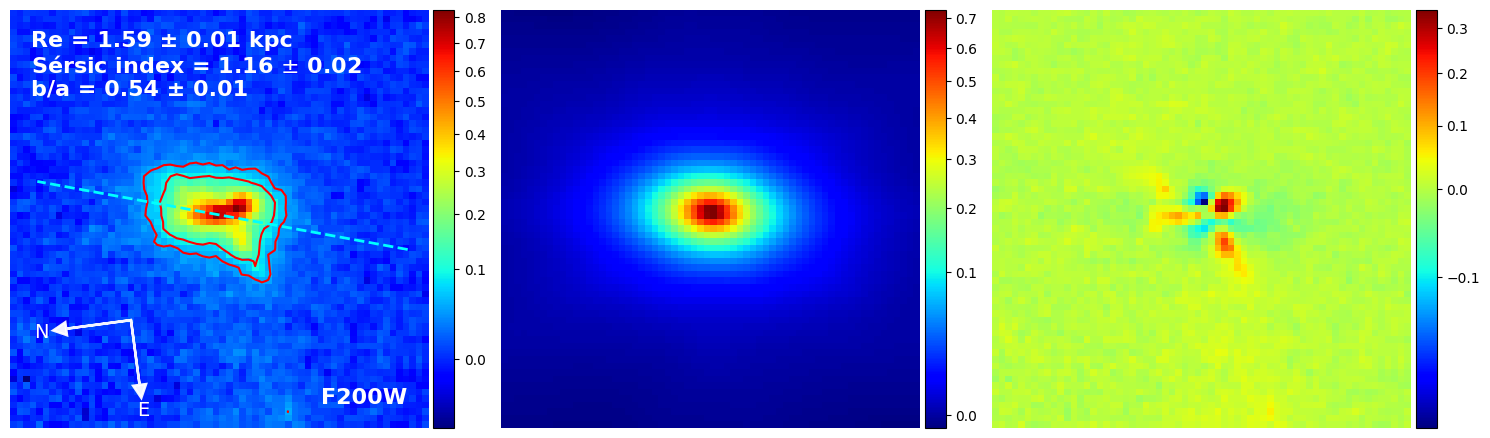}
\includegraphics[width=0.99\textwidth]{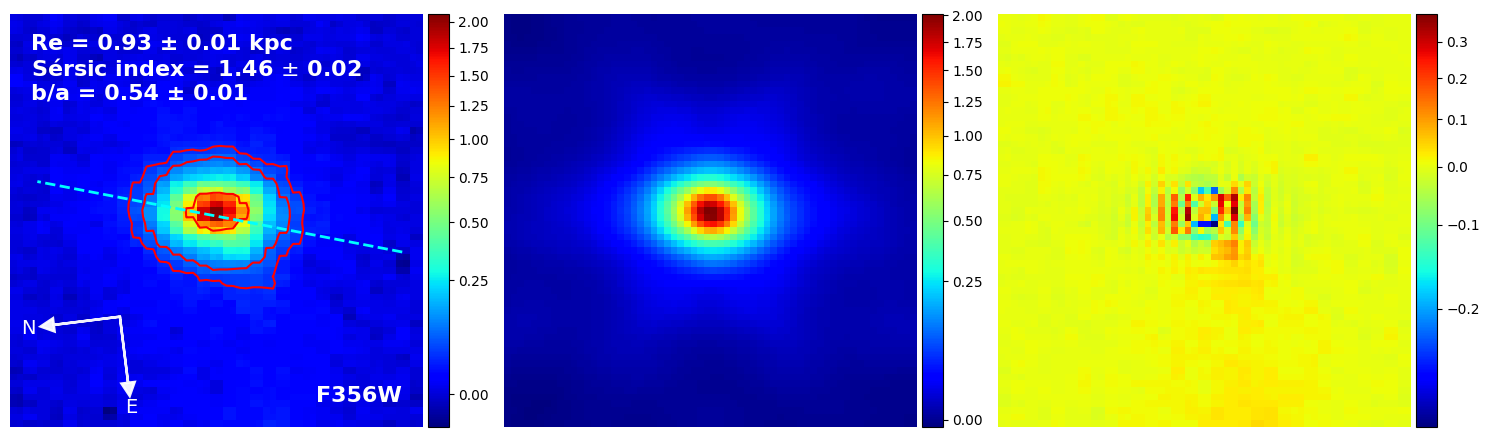}
  \caption{\small{GALFIT fitting of galaxy J0305M31-He1-2708 in F200W (upper) and F356W (lower). We use GALFIT to measure the effective radius ($R_e$), Sérsic index, axis ratio, and position angle. The cyan dashed line marks the major axis. Each row shows (left to right) the JWST image, GALFIT model, and residuals (data minus model).
}}\label{fig:J0305_g_pa}
\end{figure*}

We use the galaxy decomposition tool GALFIT \citep{peng02,peng10}\footnote{https://users.obs.carnegiescience.edu/peng/work/galfit/galfit.html} to fit the position angle (PA) and S\'ersic index of this star-forming galaxy (SFG). We note that J0305M31-He1-2708 exhibits different morphologies in the F200W and F356W bands, so we fit the F200W and F356W bands individually. The point-spread function (PSF) was generated by stacking stars in each NIRCam filter within the same field using \texttt{SourceExtractor++} \citep{bertin20}, and the model fitting was then convolved with this PSF. Prominent residuals in the F200W fit (data minus the GALFIT model) are oriented at an angle to the projected photometric major axis, suggesting a tentative outflow. The fitted PA angles are plotted in Figure \ref{fig:J0305_g_pa}. The S\'ersic indices in F200W and F356W are 1.16 $\pm$ 0.02 and 1.46 $\pm$ 0.02, respectively, indicating a tentative disk-like structure based on the disk galaxy classification criteria in \citet{sun24}. We discuss this in detail in Section \ref{sec:0305_discussion}. The effective radius ($R_e$) and $b/a$ of the galaxy in F200W are 1.59 $\pm$ 0.01 kpc and 0.54 $\pm$ 0.01, respectively, and in F356W they are 0.93 $\pm$ 0.01 kpc and 0.54 $\pm$ 0.01. We compare our measured galaxy $R_e$ in F200W and stellar mass with the empirical galaxy size (observed in the rest-frame 5000~\AA)-mass relation at $z \sim 2.7$ in \citet{wel14}, and find that our detected dusty SFG lies between early-type and late-type galaxies. 

\subsection{Comparison with other star-forming galaxies and DLA hosts at similar redshifts}


In this section, we contextualize our detected galaxies around these two VSMGII systems by comparing their properties with those of other star-forming galaxies at similar redshifts. We first compare the \lya~emission-line flux, luminosity, and $W_r$ of the two LAEs around the $z \sim 4.86$ VSMGII pair with those from other LAE surveys, DLA systems, and \MGII~hosts. In the upper panel in Figure \ref{fig:dla-lae}, we plot the \lya~line flux in the DLA host against the DLA log N({H~{\sc i}), with the colormap of velocity width ($\Delta v$) of Si~{\sc ii}. The full DLA-host sample is collected in \citet{kro17}. We extrapolate the $N$(H~{\sc i}) of our \MGII~system from the $W_r$ of \MGII($\lambda$2796) using the relation in \citet{lan18}: 
\begin{equation}
N_{\textrm{H {\sc i}}}~(\textrm{cm}^{-2}) = A \left( \frac{W_{\lambda2796}}{1 \textrm{\AA}} \right)^\alpha (1+z)^\beta
\end{equation}
where $A = 10^{18.96 \pm 0.10}$ cm$^{-2}$, $\alpha = 1.69 \pm 0.13$, and $\beta = 1.88 \pm 0.29$. Except for one exceptionally bright \lya~emission from a DLA host in \citet{kro15}, we find that the stronger column density of H~{\sc i} is, the brighter the host’s \lya~emission. Our detected \MGII~surrounding LAE is consistent with this trend. We also note that the strength of \lya~emission does not show a significant correlation with the velocity width of the absorption line.
In the lower panel of Figure \ref{fig:dla-lae}, we plot the \lya~$W_r$ from the survey in MUSE Hubble Ultra Deep \citep{hashimoto17,wiso16} and the JWST Advanced Deep Extragalactic Survey (JADES) survey \citep{wits24}. We find that the $W_r$ of our VSMGII-LAE is consistent with the median $W_r$ of LAE measured in \citet{hashimoto17} at $z\sim$ 4.9.


\begin{figure}
\resizebox{\hsize}{!}{\includegraphics{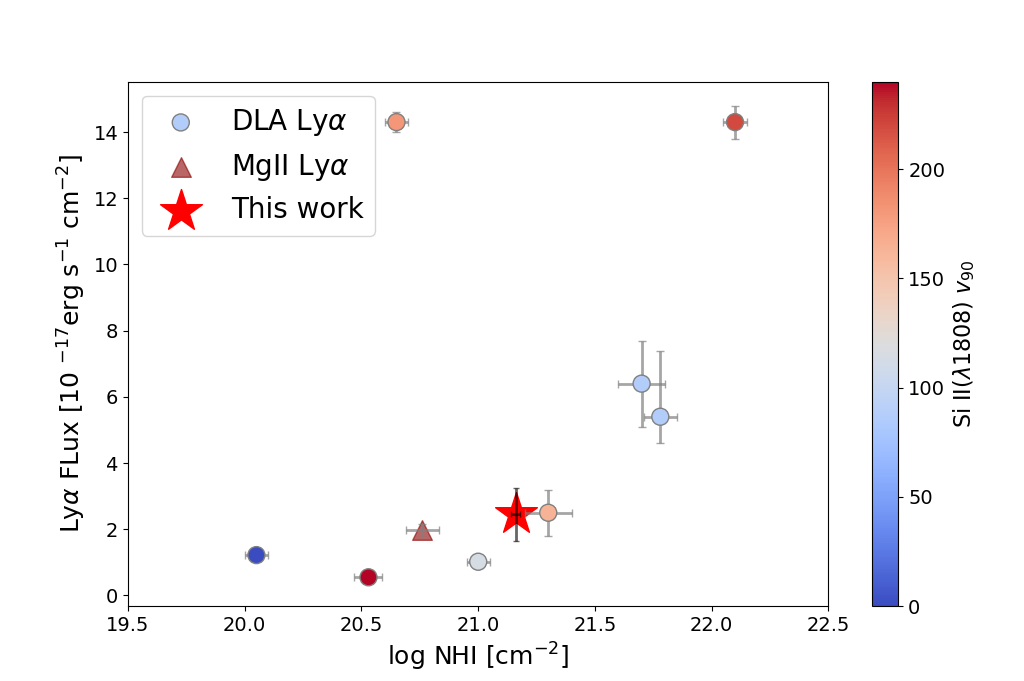}}
\resizebox{\hsize}{!}{\includegraphics{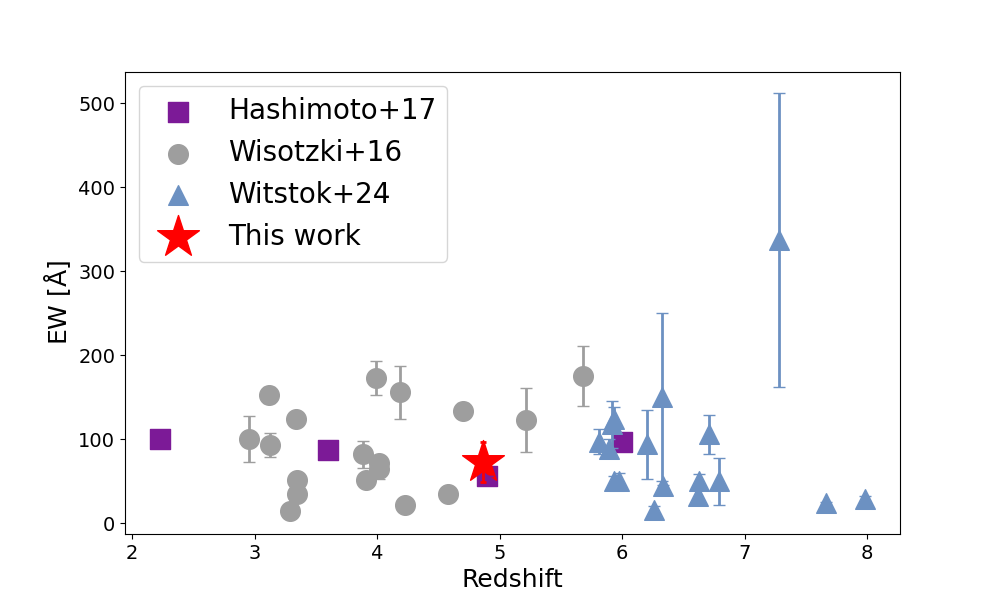}}
\caption{$Upper$: Comparison of VSMGII-\lya~line flux with other DLA-\lya~hosts. The DLA-\lya~data points are color-coded by the velocity width $v_{90}$ of the Si~{\sc ii}($\lambda$1808) line. The \MGII~system ($W_r = 2.033 \pm 0.187$ \AA~and $v_{90}(\lambda$2796) = 355 \kms) at $z = 3.188$ in the MAGG survey \citep{marta24} is plotted as a red triangle. The column density of H~{\sc i} for the two \MGII~systems, including our $z = 4.8651$ VSMGII system, is extrapolated from the $W_r$(Mg~{\sc ii})-$N$(H~{\sc i}) relation in \citet{lan18}. $Lower:$ Equivalent widths of LAEs detected in the MUSE Hubble Ultra Deep Field and the JADES survey at $2 < z < 8$.}\label{fig:dla-lae}
\end{figure}

We then compare our detected dusty SFG around the $z = 2.5661$ VSMGII with DLA hosts detected by ALMA at $z$ = 2-4 \citep{neeleman17,neeleman18,prochaska19}. The SFR of our VSMGII counterpart is consistent with the SFR of DLA submillimeter-detected hosts ($\sim$ 110 \myr). \citet{prochaska19} report a moderate SFR ($\sim$ 20 \myr) for the DLA host 
J1201+2117, which is claimed to be undergoing a major merger at $z \sim$ 3.79. 




\section{Discussion}\label{sec:discussion}



\subsection{J1306+0356}

\subsubsection{Possibility of overdensity}
We first discuss the possibility that the $z = 4.8651$ VSMGII resides in an overdense region. We estimate the random field number density ($N_{\textrm{gal}}/c\textrm{Mpc}^3$) of LAEs at $z \sim 4.8$ using the \lya~luminosity function measured from the MUSE $Hubble$ Deep data in \citet{drake17}. For log$L_{\textrm{Ly}\alpha} > 42.0$, the $\Phi(\textrm{log}L>42/c\textrm{Mpc}^3)$ is about $10^{-2}$. Given that our two LAEs have a redshift difference of $\Delta z = 0.002$, the number density within the comoving volume (MUSE FoV times $\Delta z$) is $1.14 \times 10^{-4}$ Mpc$^3$. This suggests that the two LAEs and the VSMGII gas reside in an overdense region with $\delta \sim 10^4$.

High $N$(H~{\sc i}) gas in overdense regions has been detected at similar redshifts, and suggested that the cold gas content needs to be increased in fiducial cosmological simulations. \citet{heintz24a} reported ten galaxies with strong DLA features at $z=$ 5.4--6.0 from JWST NIRSpec FRESCO (GO-18905, PI: Oesch) survey. The ten DLA-bearing galaxies reside in a foreground protocluster with more than 25 galaxies at similar redshifts ($z =$ 5.35--5.40). Such high $N$(H~{\sc i}) and nearly neutral gas content in overdense regions at the end of the reionization epoch may play a key role in cosmic assembly history and the transition phase of IGM large-scale structure. Our detection of the metal-enriched and kinematically ultra-large system may further reveal baryonic transfer processes in these overdense regions at $z \sim 5$.


\subsubsection{Metallicity and ionization state of the ultra extended system}\label{sec:cloudy}

From Figure \ref{fig:lya} we see that LAE1 exhibits an asymmetric double-peaked profile with an extension on the red side towards the absorbing system at $z$ = 4.8821. We tentatively suggest that LAE1 kinematically moves from the absorbing system at $z = 4.8634$ towards the one at $z = 4.8821$, tracing the filaments between dense and cold (probably neutral) gas in the two absorbing systems. From Figure \ref{fig:J1306_abs_complete}, we see that the two absorbing systems share similar metal species, with the bluer system ($z = 4.8821$) exhibiting slightly higher equivalent widths in the Fe~{\sc ii}, Si~{\sc ii}, and Al~{\sc ii} lines. Given that the two systems likely share a common origin and display an extremely large combined velocity width (greater than 1000 \kms), we use CLOUDY \citep{fer17} to model the ionization state, metallicity, and gas density of these systems.

Although we do not have direct measurements of $N$(H~{\sc i}), we can still derive the {\bf relative} distribution of metallicity, dust content, gas density, and ionization state between these two absorbing systems. First, we varied the stopping column density of H~{\sc i} and found that when log $N$(H~{\sc i}) $<$ 18.0, even increasing the gas metallicity to solar could not reach the detected column densities of Si~{\sc ii}, Al~{\sc ii}, and Fe~{\sc ii}. We then set log $N$(H~{\sc i}) between 20 and 22 (i.e., DLA systems) and varied the metallicity log($Z/Z_{\odot}$) between --3 and 1. The dust content was first scaled with metals and adjusted accordingly with changes in metallicity.

In all modeling results with varying metallicity, system 1 exhibits a metallicity that is $\sim$1 dex lower than that of system 2 (based on N(Zn~{\sc ii})). The ionization state parameter log $U$ of system 1 is about 0.7 dex lower than that of system 2, as indicated by the ratio of Al~{\sc ii} to Al~{\sc iii}. The ionization parameter $U$ is defined as $U = \Phi$(H)/(n(H)$\times$c), where $\Phi(H)$ is the flux of ionizing photons and c is the light speed. Specifically, when log $N$(H~{\sc i}) = 21.0, the log($Z/Z_{\odot}$) of system 1 and system 2 is --2.5 and --1.5, respectively (see Extended Data Figure \ref{fig:cloudy}). Such metallicity range is also consistent with predictions of DLA at $z\sim$ 4.8 from the statistical metallicity distributions of large DLA samples presented in \cite{raf12}. We note that Si~{\sc ii} exhibits overabundances compared to other elements, regardless of changes in metallicity and dust content. We also found that, for system 2, the modeled log N(Fe~{\sc ii}) is consistently lower than the observed value by $\sim$0.4 dex, whereas for system 1, the modeled value is higher than the observed value by about 0.4 dex. Given that iron is more sensitive to dust depletion compared to silicon and zinc \cite{jen86}, system 2 should contain more dust than system 1.

In summary, this kinematically ultra-wide ($\Delta v > 1000$ \kms) absorbing structure consists of two systems. The bluer system has lower metallicity, less dust, and is slightly more ionized than the redder one. Both systems share similar metal species and are likely connected by the \lya~emitting gas, residing in an overdense region at $z \sim 5$. We also keep the possibility that fainter host galaxies or obscured massive galaxy(ies) are located closer to the absorbing gas structure. Based on the ASPIRE-ALMA results in \citet{fengwu24}, 66 $\pm$ 7\% of star formation at $z \sim 5$ remains obscured.

\subsection{J0305--3150}\label{sec:0305_discussion}

\subsubsection{Tentative Disk?}
A DLA host harboring a rotational disk has been reported at $z \sim$ 3-5 \citep{neeleman20}. Because VSMGII systems have a high probability to be associated with a high H~{\sc i } column density, we discuss whether our host has a similar rotational disk at the cosmic noon.
We first measure the velocity gradient of the He~{\sc i} $\lambda$10830 line following the method in \citet{nelson23}. For each row of the grism 2D spectra along the dispersion direction, we fit a light-weighted center using a Gaussian profile. We then calculate the differences between the light-weighted center of the He~{\sc i} $\lambda$10830 line and the light-weighted center of the grism image. Note that the velocity derived from these two centers reflects contributions from both the spatial and kinematic properties of the line. Since we do not have multiple medium-band images to differentiate these two contributions, we can only constrain the $v_{\rm rot}$ limit.

The observed velocity of He~{\sc i} line is then calculated as $v_{\rm obs} = (v_{\rm max} - v_{\rm min}) / 2$ from the velocity map. The $v_{\rm max}$ and $v_{\rm min}$ represent the maximum and minimum velocities measured from the observed velocity field in the 2D grism data. For He~{\sc i} line, we find $v_{\rm obs} = 248 \pm 52$ \kms. The rotation velocity, $v_{\rm rot}$, is derived from $v_{\rm rot} = v_{\rm obs} / \sin i$, where $i$ is the inclination angle. As mentioned in \citet{nelson23}, the intrinsic $q_0$ of the galaxy cannot be determined, so we assume $0.05 < q_0 < q$ to estimate the uncertainties in $v_{\rm rot}$, where $q$ is the observed major-to-minor axis ratio. We note that the velocity dispersion is contributed by both the instrument and the target. NIRCam has a resolution of $R \sim 1550$ at $\lambda \sim 3.5 \mu$m, corresponding to $\sim 84$ \kms. Considering the instrumental effect, the corrected $v_{\rm rot}$ for the galaxy and the emission line should be below 352 \kms.

We apply a dynamical modeling approach to the He~{\sc i} line to further assess $v_{\rm rot}$, following the method described in \citet{li23}. The velocity map is obtained using a similar procedure as above, followed by Markov chain Monte Carlo (MCMC) sampling to derive the best-fit kinematic parameters and their uncertainties. This method builds on the approach developed by \citet{out20}. First, we fit the 2D surface brightness in the source plane and use the derived morphological parameters to generate the velocity field. We then use the F356W imaging to construct the velocity dispersion field, assuming an intrinsically circular disk. The fitting results are presented in Figure~\ref{fig:J0305_v_gradient}. For the He~{\sc i} line, we derive $v_{\rm rot} = 115^{+97}_{-86}$~\kms, indicating a tentative rotating disk.


%



\begin{figure*}
  \includegraphics[width=0.85\textwidth]{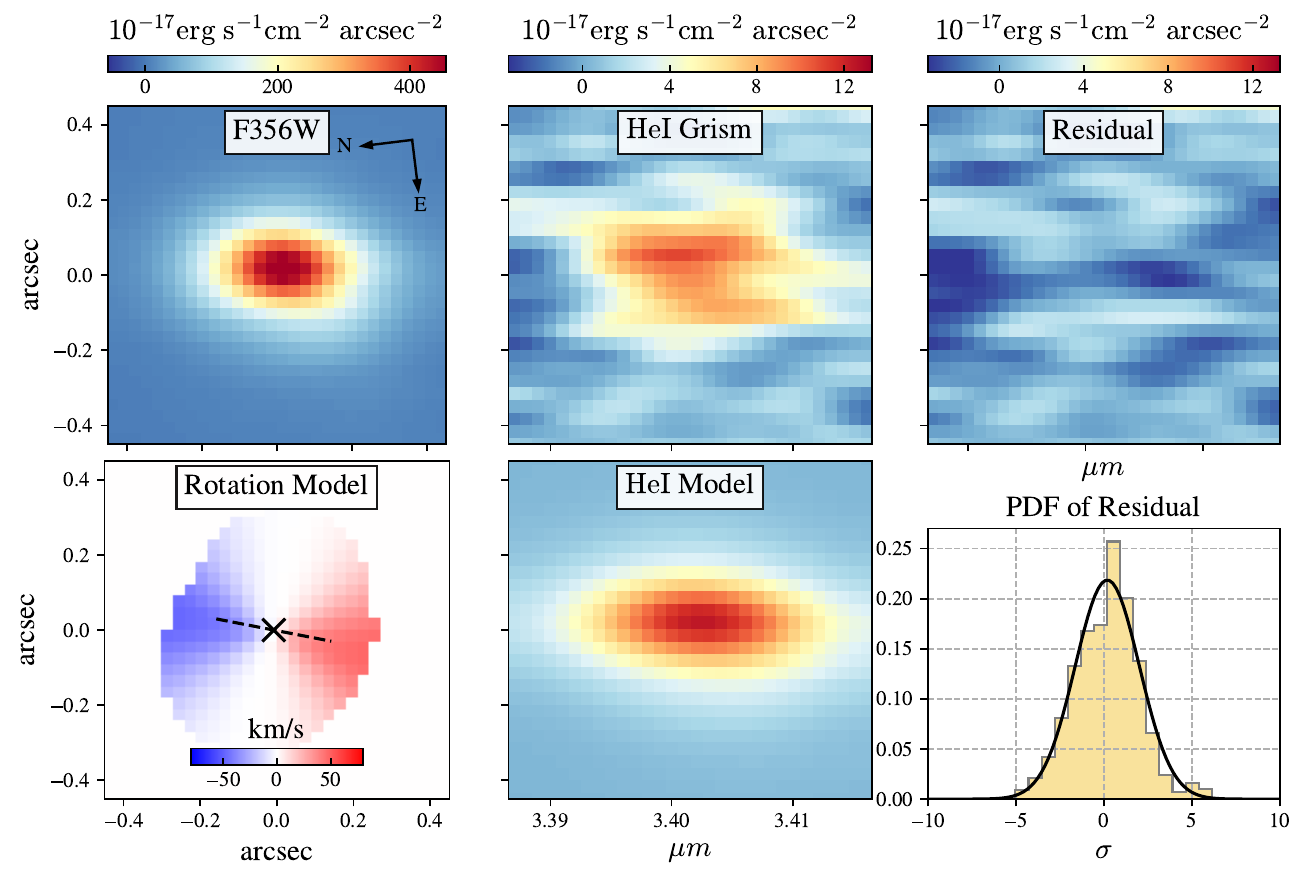}
  \caption{\small{Dynamical modeling for the He~{\sc i} line in J0305M31-He1-2708. We plot the galaxy’s F356W image and the He~{\sc i} grism spectra in the upper panel. In the lower panel, we show the rotation model of the line and the best-fit S\'ersic model convolved with the point spread function, following the method in \citet{li23}. In the last column of panels, we present the fitting residuals.
}}\label{fig:J0305_v_gradient}
\end{figure*}

\subsubsection{Comparison with cosmological simulation }

To further understand the kinematics and dynamics of metal-enriched cold gas and its role in disk and galaxy formation, we compare the H~{\sc i} gas distribution around similar disk-like galaxies (in terms of SFR, $v_{\rm rot}$, $M_*$, and $R_e$) in the cosmological simulation IllustrisTNG100 \citep[hereafter TNG-100;][]{nelson18}. Disk-like galaxies are identified by requiring a disk mass fraction, $f_{\rm disk}$, greater than 0.5. The $f_{\rm disk}$ values are obtained using the morphological decomposition method from \cite{liang24}. We select 18 disk-like galaxies and find that high N(H~{\sc i}) gas within 50 kpc of the host galaxies shows a high incidence along both the major and minor axes, i.e., it does not purely trace inflows or outflows. The angular momentum of the high N(H~{\sc i}) gas is not perpendicular with the disk angular momentum, indicating dynamical complexity beyond simple inflow or outflow models. We present one example in Figure \ref{fig:simulation}. 
We plot the column density distribution of H~{\sc i} gas and the angular momentum directions of the disk and H~{\sc i} gas. The H~{\sc i} gas is derived from the total hydrogen content in TNG100 using a post-processing model based on \citet{gne11}. 

From the simulation, we see that high density H~{\sc i} gas penetrates the hot CGM and then accretes onto the galaxy. Analytical models predict that the mode of cold gas accretion onto galaxies evolves with redshift. The fraction of cold-in-hot mode increases with redshift in a large-mass halo ($M_h > 10^{12}M_{\odot}$) at $z > 1$ \citep{dekel06}. The penetrated cold flow further sustains star and disk formation. Meanwhile, the turbulence and pressure instability in the inner CGM caused by cold flows enhance the probability of galactic outflows escaping from the galaxy. Therefore, the cold flow in high-mass halos plays a critical role in the complete virialization of the CGM, and consequently, in the transition of galaxies from star-forming to quiescent.

We estimate the halo mass of our detected SFG using the stellar-halo mass relation in \citet{gir20} and obtain log $(M_h/M_{\odot}) = 12.27^{+0.02}_{-0.60}$. The virial radius is then calculated using the relation $R_{\text{vir}} = \left( M_h / \left( (3/4)\pi \times 200 \times \rho_{\text{crit}} \right) \right)^{1/3}$, which gives $R_{\text{vir}} = 245$ kpc. Our detected VSMGII gas resides in the 0.16 $R_{vir}$. Combining the host galaxy's outflow-like features from JWST F200W and F356W imaging, its classification as not yet early-type, and the tentative rotating disk structure, we suggest that the high velocity VSMGII gas is accreting onto the galaxy, supporting starburst and disk formation, while also being affected by the tentative galactic outflow given from the JWST (in particular F200W) imaging morphology fitting.

To test whether the cold CGM angular momentum is corotating with the disk-like host galaxy or significantly shaped by outflows, we need deep IFU observations that deliver three-dimensional maps of CGM emission and spatially resolved kinematics for direct comparison with the galaxy rotation field.

\begin{figure*}
  \resizebox{\hsize}{!}{\includegraphics{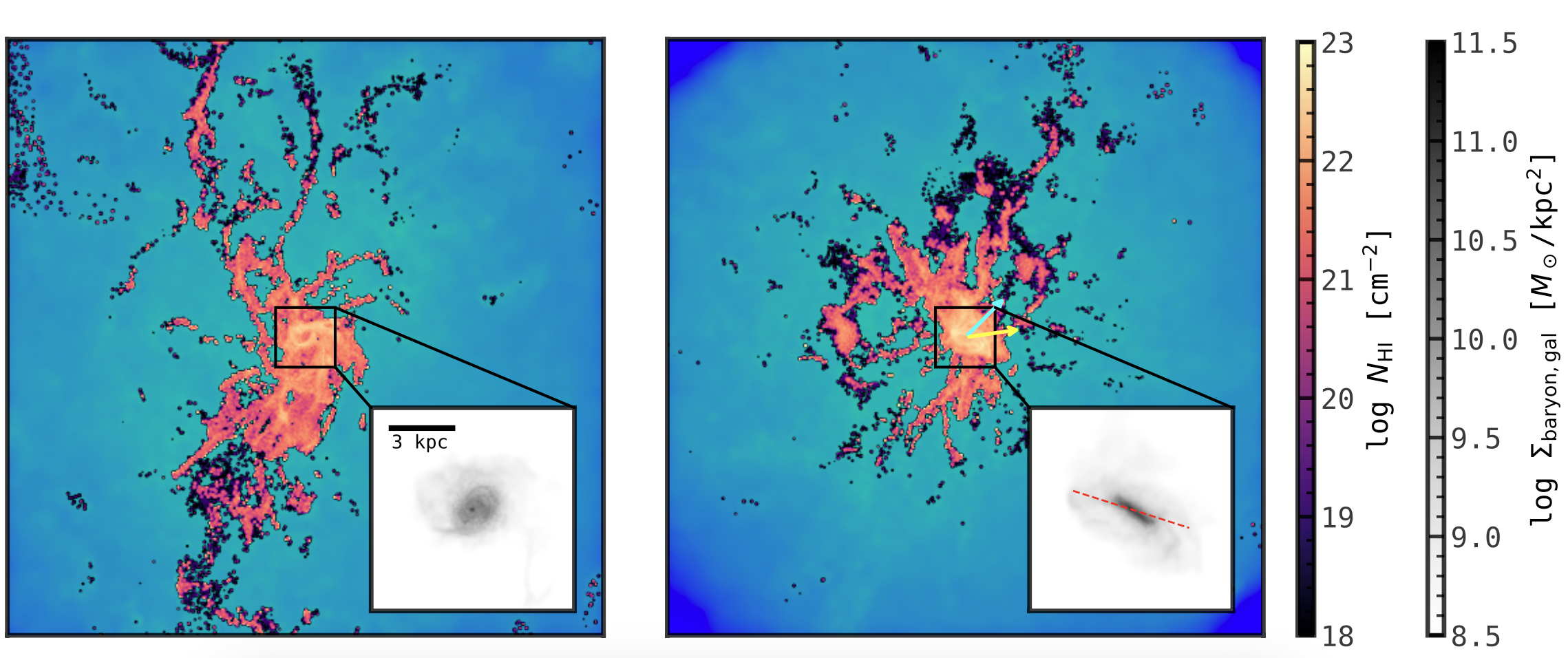}}
  \caption{\small{Example of the H~{\sc i} gas column density distribution around one disk galaxy at $z \sim 3$ from TNG100, selected based on the disk-like galaxy criteria in \citet{liang24}. The simulated galaxy has similar stellar mass, halo mass, SFR, and $R_e$ to our detected galaxy. The left and right panels show face-on and edge-on views of the same galaxy on a 100 $\times$ 100 kpc scale, respectively. The zoom-in box shows the stellar+gas content of the galaxy on a 10 $\times$ 10 kpc scale. The red dashed line is the major axis of the galaxy. The blue and yellow arrows in the right panel indicate the angular momentum directions of the disk and H~{\sc i} gas, respectively.
}}\label{fig:simulation}
\end{figure*}







\section{Summary}\label{sec:summary}

We present two ultra-strong and kinematically broad \MGII~absorbing systems that trace disturbed cold gas and their associated galaxies at $z > 2$. We search for VSMGII absorber–galaxy counterparts using VLT/MUSE, JWST, HST, and ALMA data. One system is an ultra-strong, metal-enriched absorber (comprising two VSMGIIs) with a velocity width exceeding 1000~\kms, accompanied by two surrounding LAEs at an impact parameter of $\sim$200~kpc. The two absorbing systems exhibit interactions in metals, dust, and ionization, and are possibly connected by \lya~emitting gas. The other VSMGII system, at $z \sim 2.5$, resides in the CGM ($D = 38$~kpc) of a massive, dusty star-forming galaxy, likely hosting a tentative rotating disk.

We find that the VSMGII-traced cold gas resides in complex environments, both kinematically and dynamically. This cold gas plausibly influences disk formation and evolution, as well as the large-scale structure at high redshift. Although the two systems alone cannot provide strong, definitive conclusions on whether the cold gas traces inflows, outflows, or disk formation, they shed light on the role of cold gas in feeding early disk growth and in driving metal and dust enrichment in early overdense regions, particularly with the advent of ground-based IFU instruments and the JWST. Future observations of obscured and faint star-forming galaxies will further explore the connection between these processes and the cold gas–galaxy kinematics, dynamics, and metallicity.

\acknowledgements
We thank the anonymous referee for the careful reading and constructive suggestions. We thank fruitful discussions with Weichen Wang, Valentina D'Odorico, Fengwu Sun and Pasquier Noterdaeme. SZ acknowledges support from the National Science Foundation of China (no. 12303011). JBC acknowledges funding from the JWST Arizona/Steward Postdoc in Early galaxies and Reionization (JASPER) Scholar contract at the University of Arizona. WS and LCH acknowledge support from the National Key R\&D Program of China (2022YFF0503401), the National Science Foundation of China (11991052, 12233001), and the China Manned Space Project (CMS-CSST-2021-A04, CMS-CSST-2021-A06). PPJ is supported in part by contract ANR-22-CE31-0009 (HZ-3D-MAP) from the Agence Nationale de la Recherche.

This work is based on observations made with the Magellan/FIRE, NASA/ESA/CSA James Webb Space Telescope and Hubble Space Telescope, VLT/MUSE, and ALMA telescopes. The JWST data were obtained from the Mikulski Archive for Space Telescopes at the Space Telescope Science Institute, which is operated by the Association of Universities for Research in Astronomy, Inc., under NASA contract NAS 5-03127 for JWST. These observations are associated with program \#2078. Support for program \#2078 was provided by NASA through a grant from the Space Telescope Science Institute, which is operated by the Association of Universities for Research in Astronomy, Inc., under NASA contract NAS 5-03127. This paper makes use of the following ALMA data: ADS/JAO.ALMA\#2022.1.01077.L. ALMA is a partnership of ESO (representing its member states), NSF (USA) and NINS (Japan), together with NRC (Canada), NSTC and ASIAA (Taiwan), and KASI (Republic of Korea), in cooperation with the Republic of Chile. The Joint ALMA Observatory is operated by ESO, AUI/NRAO and NAOJ. The National Radio Astronomy Observatory is a facility of the National Science Foundation operated under cooperative agreement by Associated Universities, Inc.

Some of the data presented in this paper were obtained from the Mikulski Archive for Space Telescopes (MAST) at the Space Telescope Science Institute. The specific observations analyzed can be accessed via DOI: 10.17909/vt74-kd84.

\facilities{VLT/MUSE}
\facilities{JWST/NIRCam}
\facilities{ALMA}
\facilities{HST}

\clearpage
\bibliographystyle{aasjournal}
\bibliography{./ms}

\appendix
\restartappendixnumbering

\section{Figures}

\begin{figure*}
\resizebox{\hsize}{!}{\includegraphics{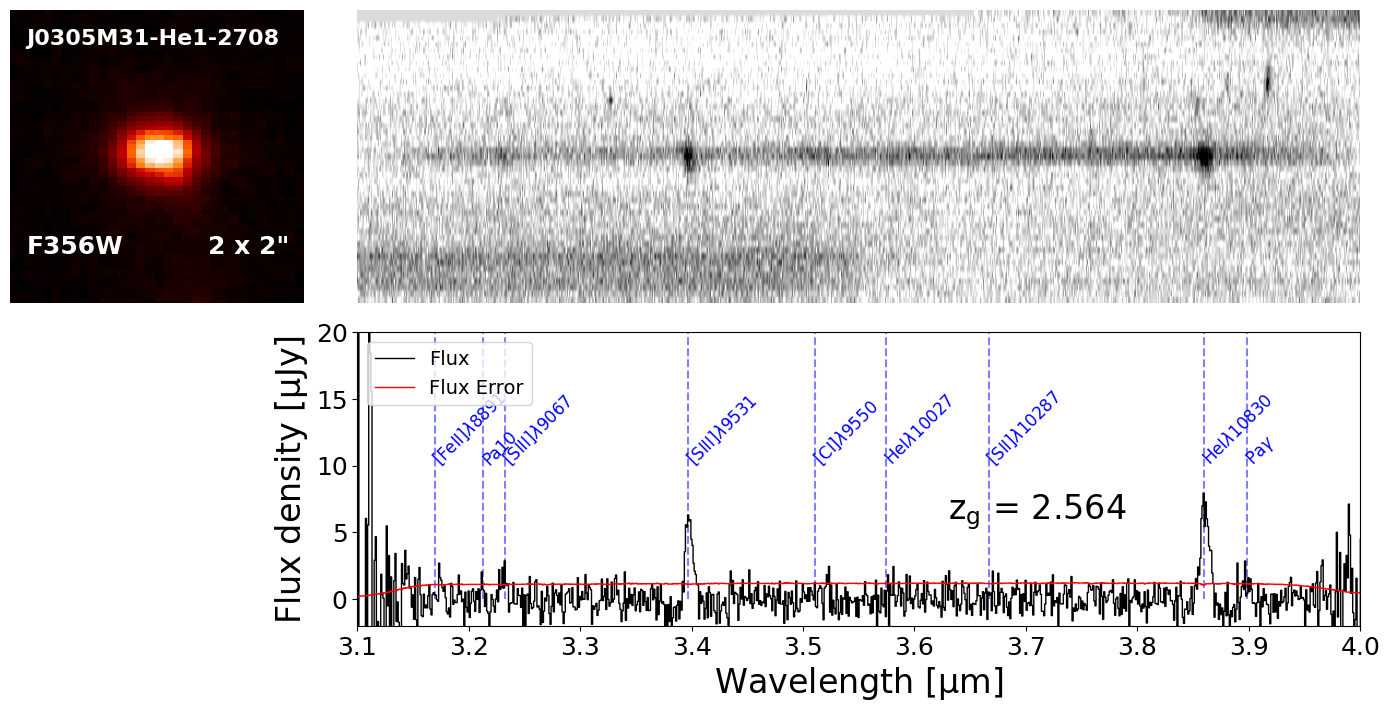}}
  \caption{\small{The 2D and 1D spectra of the host galaxy of the VSMGII absorber. The cutout imaging size is 2 $\times$ 2 $\arcsec$.
}}\label{fig:J0305_emi}
\end{figure*}

\begin{figure*}
  \resizebox{\hsize}{!}{\includegraphics{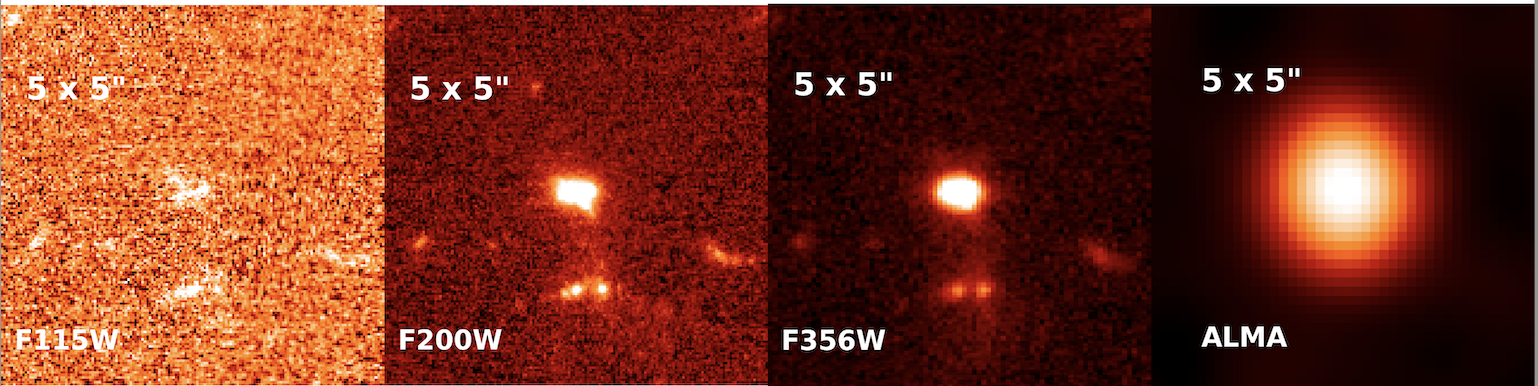}}
  \caption{\small{The JWST/NIRCam imaging of J0305M31-He1-2708 in the F115W, F200W, and F356W bands. The last panel shows the ALMA tapered imaging to increase the SNR. All images are 5 $\times$ 5$\arcsec$ in size.
}}\label{fig:J0305_jwst_alma}
\end{figure*}



\begin{figure*}
  \centering
  \includegraphics[width=0.8\textwidth]{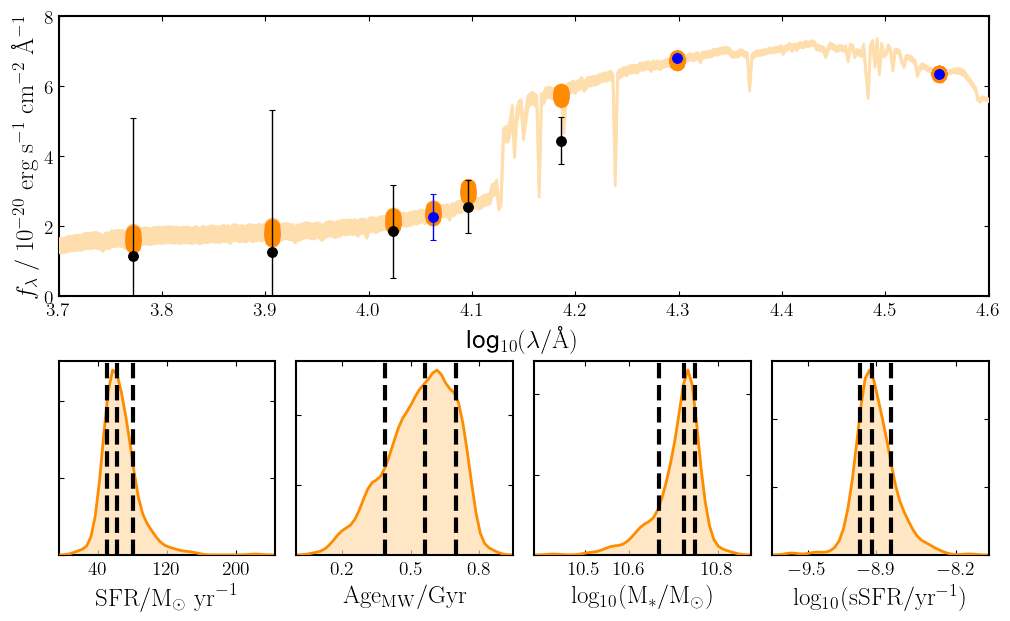} 
  \caption{\small{Spectral energy distribution fitting of J0305M31-He1-2708 using BAGPIPES. We use photometry from JWST/NIRCam F115W, F200W, and F356W bands, and from HST F105W, F125W, F160W, F606W, and F814W bands, shown as blue and black points in the upper panel, respectively. The orange ovals and curve in the upper panel represents the modelled photometry in each observed band and the} Spectral Energy Distribution. The lower panel shows the likelihood distributions of the fitted SFR, galaxy age, stellar mass, and specific SFR, respectively. The black dashed lines mark the 16th, 50th, and 84th percentiles of the posterior. }
  \label{fig:J0305_sed}
\end{figure*}

\begin{figure*}
  \resizebox{0.9\hsize}{!}{\includegraphics{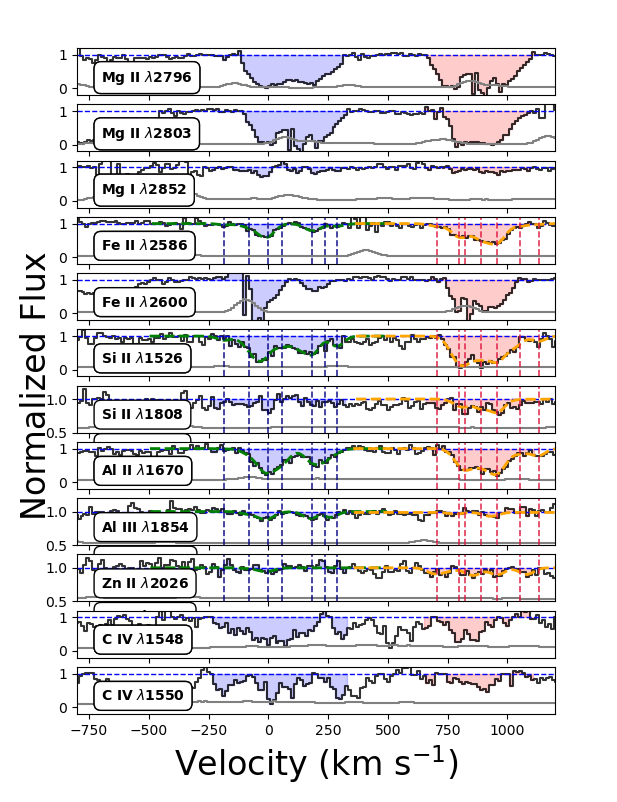}}
  \caption{\small{Complete velocity profile of absorption lines of the two USMg~{\sc ii} structures towards J1306+3150 at $z = 4.87$. System 1 at $z = 4.8634$ and system 2 at $z = 4.8821$ are plotted in blue and red shadow, respectively. The Voigt fitting profiles of Fe~{\sc ii}, Si~{\sc ii}, Al~{\sc ii}, Al~{\sc iii}, and Zn~{\sc ii} lines in these two systems are plotted with green and orange dashed lines, respectively. The vertical dashed lines indicate the positions of the subcomponents in the fitting of the two systems. The quasar spectra were taken with Magellan/FIRE with a spectral resolution of $R \sim 6000$.
}}\label{fig:J1306_abs_complete}
\end{figure*}

\begin{figure*}
  \resizebox{1.0\hsize}{!}{\includegraphics{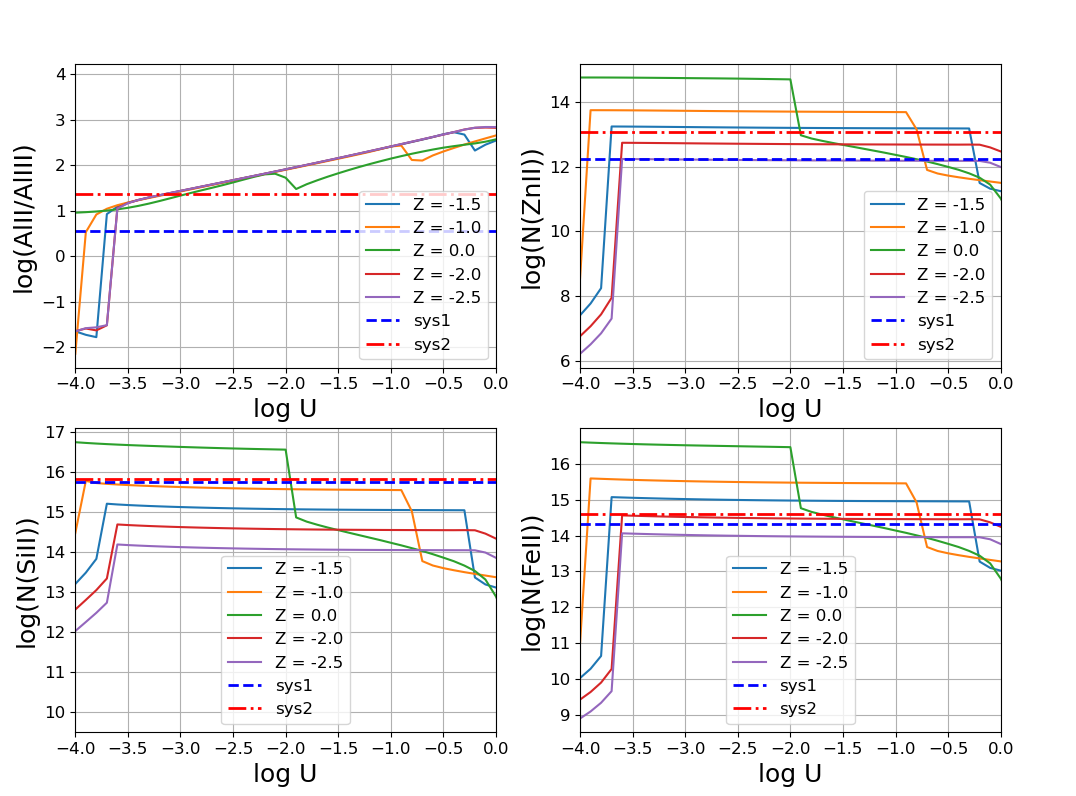}}
  \caption{\small{CLOUDY modeling of two absorbing systems at $z \sim 4.9$ along the sightline J1306+3150. The observed values of system 1 at $z = 4.8634$ and system 2 at $z = 4.8821$ are plotted as blue dashed and red dash-dotted lines, respectively. The modeling curves are plotted with different metallicities (log($Z/Z_{\odot}$)) ranging from --2.5 to 0. From the upper left to the lower right, we plot the column density ratio of Al~{\sc ii} to Al~{\sc iii} against ionization parameter, and the column densities of Zn~{\sc ii}, Si~{\sc ii}, and Fe~{\sc ii} against ionization parameter, respectively.
}}\label{fig:cloudy}
\end{figure*}

\begin{table}
\centering
  \caption{Re-measurements of the ion equivalent widths ($W_r$) of J1306+1356 are presented. We denote the system at $z = 4.8634$ as system 1 and the one at $z = 4.8821$ as system 2. The blended region refers to Mg~{\sc ii}($\lambda$2803) in system 1, which is blended with Mg~{\sc ii}($\lambda$2796) in system 2. Columns 4 and 5 show the Voigt-profile fitted column densities of ions with a Doppler parameter $b = 25$ \kms. The uncertainties are given by varying the $b$ value by $\pm$ 5 \kms. We do not measure the column densities of \MGII~and \CIV~lines due to strong blending in these two systems. \label{table:absorption}}
\begin{tabular}{lllll}
\hline
Ions                       & $W_r$(System 1)  & $W_r$(System 2)                    & log N (System1)   &  log N (System2) \\  
                           & (\AA)           & (\AA)       &  (cm$^{-2}$)             &  (cm$^{-2}$)                     \\ 
\hline              
Mg~{\sc ii}($\lambda$2796)   & 2.891 $\pm$ 0.109     &  -                  &   -            &   -           \\ 
Mg~{\sc ii} blended          & \multicolumn{2}{c}{2.959 $\pm$ 0.160}         &           -    &      -        \\ 
Mg~{\sc ii}($\lambda$2803)   &   -                 &   2.371 $\pm$ 0.097     &         -      &      -         \\ 
Mg~{\sc i}($\lambda$2852)    &  0.391 $\pm$ 0.250    &    0.359 $\pm$ 0.181  &  $<$ 13.98             &       $<13.94$        \\
Fe~{\sc ii}($\lambda$2586)   &  0.553 $\pm$ 0.141    &   1.141 $\pm$ 0.123   &   14.32 $\pm$ 0.02             &    14.60 $\pm$ 0.12          \\
Si~{\sc ii}($\lambda$1526)   &  0.915 $\pm$ 0.062    &   1.137 $\pm$ 0.060   &   15.76 $\pm$ 0.03              &   15.82 $\pm$ 0.05            \\
Al~{\sc ii}($\lambda$1670)   &  0.923 $\pm$ 0.069    &   0.988 $\pm$ 0.067   &   13.59 $\pm$ 0.11             &    13.726 $\pm$ 0.10           \\
Al~{\sc iii}($\lambda$1854)  &  0.165 $\pm$ 0.108    &   0.106 $\pm$ 0.105   &   13.034 $\pm$ 0.01            &      13.649 $\pm$ 0.11       \\
Zn~{\sc ii}($\lambda$2026)   &  $<$ 0.027    &   0.160 $\pm$ 0.092                    &   $<$12.237             &    13.056 $\pm$ 0.05         \\
C~{\sc iv}($\lambda$1548)    &  1.141 $\pm$ 0.298   &    0.552 $\pm$ 0.267   &        -       &         -      \\ 
C~{\sc iv}($\lambda$1550)    &  0.791 $\pm$ 0.358  &     0.435 $\pm$ 0.251   &           -    &         -      \\ 
\hline
\end{tabular}
\end{table}

\end{document}